\documentclass[a4paper]{PoS}
\usepackage{wrapfig}
\usepackage{graphicx}
\usepackage{amsmath,amssymb,amsfonts,graphics,graphicx,amscd,amsfonts,epsfig,color}
\usepackage{dsfont}
\usepackage{subfig}
\usepackage{cite}
\usepackage{enumitem}
\usepackage{here}
\newcommand{\rf} [1] {(\ref{#1})}
\newcommand{\vev}[1] {{\ensuremath{\langle {#1} \rangle}}}
\newcommand{\mf}      {\ensuremath{m_{\rm f}} }
\newcommand{\epsilonm}{\ensuremath{\epsilon} }
\newcommand{\Sb}      {\ensuremath{S_{\rm b}} }
\newcommand{\Sf}      {\ensuremath{S_{\rm f}} }
\newcommand{\detM}    {\ensuremath{\det{\cal M}}}
\newcommand{\tr}      {\mathrm{tr}}
\newcommand{\Tr}      {\mathrm{Tr}}
\newcommand{\nn}      {\nonumber}
\usepackage[colorinlistoftodos,shadow]{todonotes}

\title{Dynamical compactification of extra dimensions in the Euclidean type IIB matrix model: A numerical study using the complex Langevin method}

\ShortTitle{Dynamical compactification of extra dimensions}

\author{Konstantinos N. Anagnostopoulos$^{a}$, Takehiro Azuma$^{b}$, Yuta Ito$^{c}$, Jun Nishimura$^{cd}$ and \speaker{Stratos Kovalkov Papadoudis}$^{a}$\\
		\\
		\llap{$^{a}$}Physics Department, National Technical University,\\
		Zografou Campus, GR-15780 Athens, Greece\\
		\llap{$^{b}$}Institute for Fundamental Sciences, Setsunan University,\\
		17-8 Ikeda Nakamachi, Neyagawa, Osaka, 572-8508, Japan\\
		\llap{$^{c}$}KEK Theory Center, High Energy Accelerator Research Organization,\\
		1-1 Oho, Tsukuba, Ibaraki 305-0801, Japan\\
		\llap{$^{d}$}Graduate University for Advanced Studies (SOKENDAI),\\
		1-1 Oho, Tsukuba, Ibaraki 305-0801, Japan\\
		\\
        E-mails: \email{konstant@mail.ntua.gr}, \email{azuma@mpg.setsunan.ac.jp}, \email{yito@post.kek.jp}, \email{jnishi@post.kek.jp} and \email{sp10018@central.ntua.gr}
		\\
}

\abstract{The type IIB matrix model is conjectured to be a
  nonperturbative definition of type IIB superstring theory. In this
  model, spacetime is a dynamical quantity and compactification of
  extra dimensions can be realized via spontaneous symmetry breaking
  (SSB). In this work, we consider a simpler, related, six dimensional model in its
  Euclidean version and study it numerically. Our calculations provide
  evidence that the SO(6) rotational symmetry of the model breaks down
  to SO(3), which means that the theory lives in a vacuum where 3 out of the 6 dimensions are large
  compared to the other 3. Our results show the same SSB pattern
  predicted by the Gaussian expansion method and that they are in quantitative
  agreement. The Monte Carlo simulations are hindered by a
  severe complex action problem which is addressed by applying the
  complex Langevin method. }

\FullConference{Corfu Summer Institute 2018 "School and Workshops on Elementary Particle Physics and Gravity"\\
		(CORFU2018)\\
		31 August - 28 September, 2018\\
		Corfu, Greece}
\begin{document}
% %%%%%%%%%%%%%%%%%%%%%%%%%%%%%%%%%%%%%%%%%%%%%%%%%%%%%%%%%%%%%%%%%%%%%%%%%%%%%%%%%%%%%%%%%%
% %%%%%%%%%%%%%%%%%%%%%%%%%%%%%%%%%%%%%%%%%%%%%%%%%%%%%%%%%%%%%%%%%%%%%%%%%%%%%%%%%%%%%%%%%%
\section{Introduction}
\label{sec:intro}
% %%%%%%%%%%%%%%%%%%%%%%%%%%%%%%%%%%%%%%%%%%%%%%%%%%%%%%%%%%%%%%%%%%%%%%%%%%%%%%%%%%%%%%%%%%
% %%%%%%%%%%%%%%%%%%%%%%%%%%%%%%%%%%%%%%%%%%%%%%%%%%%%%%%%%%%%%%%%%%%%%%%%%%%%%%%%%%%%%%%%%%
Superstring theory is the most promising fundamental theory for the
unification of all interactions, including gravity. The theory is defined in
10 spacetime dimensions and the connection to the real world, where 4
dimensions are macroscopic, is realized via compactification of the
extra dimensions. This requires the introduction of many arbitrary
parameters in the theory, leading to the problem of the string
landscape. The IKKT or IIB matrix model\cite{9612115}, formally
obtained by the dimensional reduction of ten-dimensional ${\cal N}=1$
super-Yang-Mills theory to zero dimensions, is conjectured to be a
non-perturbative definition of superstring theory in the large-$N$
limit of the size of the matrices $N$. In this model, spacetime emerges
dynamically from the eigenvalues of the bosonic degrees of
freedom. Therefore, the scenario of the dynamical compactification of
extra dimensions becomes possible. Monte Carlo simulations and
analytic calculations using the Gaussian Expansion Method (GEM),
provide evidence that dynamical compactification of extra dimensions
occurs via Spontaneous Symmetry Breaking (SSB) of the rotational
symmetry of space. Monte Carlo simulations
\cite{Kim:2011cr,Aoki:2019tby,Nishimura:2019qal} provide evidence that
an infinite time emerges dynamically and 3 dimensional space
undergoes expansion leaving the remaining 6 space dimensions small,
showing that the model may realize phenomenologically interesting
cosmology. The Euclidean version of the IIB model, obtained after a
Wick rotation of the temporal dimension, has been studied
using the GEM, providing evidence that dynamical compactification
occurs via SSB of the SO(10) rotational symmetry down to
SO(3)\cite{Nishimura:2011xy}.

The Monte Carlo simulation of the Euclidean IIB matrix model, referred
to simply as the IIB model in the following, can be thought of being
the analog of lattice QCD simulations for superstring theory. Early attempts
include the simulation of simpler, related models in lower dimensions
or effective models that are thought to capture the central
dynamical properties of the model that result in the SSB of the rotational
symmetry \cite{Krauth:1998xh,Hotta:1998en,Ambjorn:2000bf,Ambjorn:2000dx,Ambjorn:2001xs}. 
The simulation of the IIB model is hindered by the
complex action problem. The effective action which results after the
integration of the fermionic degrees of freedom is complex, and it has
been conjectured that the wild fluctuations of its complex phase is
the reason that causes the SSB \cite{Nishimura:2000ds}. Configurations
which are close to being lower dimensional result in milder
fluctuations of the complex phase, therefore making them dominant in the path
integral. This effect has been examined in 
\cite{Nishimura:2000wf,Anagnostopoulos:2001yb,Nishimura:2001sq,Anagnostopoulos:2010ux,Anagnostopoulos:2011cn,Anagnostopoulos:2013xga,Anagnostopoulos:2015gua}. It
should be noted that the absence of the complex phase, like in the
$D=4$ IIB model \cite{Ambjorn:2000bf,Ambjorn:2001xs} or the phase
quenched models \cite{Anagnostopoulos:2013xga,Anagnostopoulos:2015gua}
implies absence of SSB. The
complex action problem was addressed using a reweighting-based method
\cite{Anagnostopoulos:2001yb}, but it turned out to be very hard to
determine the pattern of the SSB \cite{Anagnostopoulos:2015gua}. The complex Langevin
method (CLM) \cite{Parisi:1984cs,Klauder:1983sp} has been applied
successfully in several models with the complex action problem. CLM
defines a stochastic process which can be used to calculate the
expectation values of the observables. It is computationally simple,
but it has the disadvantage of leading to wrong results in several
known cases. Recent work \cite{Aarts:2009uq,Aarts:2011ax,Nishimura:2015pba,Nagata:2016vkn,Salcedo:2016kyy,Aarts:2017vrv} has
clarified the conditions that are necessary and sufficient for justifying the CLM
and has provided new techniques that make possible to meet
these conditions for a larger space of
parameters\cite{Seiler:2012wz,Nagata:2015uga,Tsutsui:2015tua,Nagata:2016alq,Ito:2016efb,Bloch:2017ods,Doi:2017gmk,Bloch:2017sex}. The
CLM has been recently applied  \cite{Ito:2016efb} to a simple matrix
model with severe complex action problem\cite{Nishimura:2001sq} which
has SO(4) rotational symmetry that 
is expected to spontaneously break down to SO(2)
\cite{Nishimura:2004ts}. By deforming the original model, the singular
drift problem \cite{Nishimura:2015pba} was avoided by the resulting
shift of the
eigenvalues of the Dirac operator away from zero
\cite{Mollgaard:2013qra}. By extrapolating back to the original model,
it was possible to reproduce the results of the GEM
\cite{Nishimura:2004ts}. A similar problem occurs in many interesting
problems with a complex fermionic effective action, like finite
density QCD at low temperatures.

In this talk, we present the results in \cite{Anagnostopoulos:2017gos}
where the work in \cite{Ito:2016efb} is extended to the 6D version of
the IIB matrix model. This model suffers from a severe complex action
problem due to a complex determinant appearing in the effective action
after the integration of the fermionic degrees of freedom. GEM
calculations provide evidence that the complex phase of the effective
action causes SSB of the SO(6) symmetry down to SO(3)
\cite{Aoyama:2010ry}. Using CLM and the methods employed in
\cite{Ito:2016efb}, we were able to reproduce the pattern of the SSB
which was only marginally possible by using a reweighting-based method
\cite{Anagnostopoulos:2013xga}. SSB is probed by perturbing the model
with explicit SO(6) symmetry breaking operators $\vev{\lambda_\mu}$ representing the
extent of spacetime in each direction, with the magnitude of the
perturbation controlled by a parameter $\epsilonm$. $\epsilonm$ is
later extrapolated to 0 {\it after} taking the large-$N$
limit. The singular drift problem is addressed by deforming the model
with a fermionic operator with deformation parameter $\mf$. For finite
$\mf$, the distribution of the eigenvalues is shifted away from 0 and
the singular drift problem is avoided. This is checked directly by
computing the eigenvalue distribution for small matrices, but it can also
be easily checked for all of our measurements by applying a simple
criterion proposed in \cite{Nagata:2016vkn}. In \cite{Nagata:2016vkn},
it was shown 
that the singular drift problem does not appear when
the distribution of the magnitude of the drift $u$ is suppressed
exponentially or faster  for large values of $u$.

In order to obtain the SSB pattern, a careful extrapolation to the
original model must be taken. First the large-$N$ limit is obtained
for finite $\mf$ and $\epsilonm$ values. Then, the limit 
$\epsilonm\to 0$ is taken in order to determine the SSB pattern for
a given value of $\mf$. For finite $\mf$,
SO(6) is explicitly broken down to SO(5). This is not a problem,
however, because we are looking for SSB to
SO($d$) for $d<5$. Finally the $\mf=0$ extrapolation is taken and
we find SSB to SO(3), with results that are quantitatively consistent
with GEM. 

These methods are currently applied to the original $D=10$ dimensional
IIB model\cite{future}. In this case, GEM predicts that SO(10) is broken down to
SO(3) instead of the desired SO(4) for a four dimensional
spacetime\cite{Nishimura:2011xy} and a first principle calculation is
desired. The success of the deformation method in the IIB model is
encouraging attempts to apply it to other physically interesting
systems with severe complex action problems, like in finite density
QCD \cite{Nagata:2017pgc}.

% %%%%%%%%%%%%%%%%%%%%%%%%%%%%%%%%%%%%%%%%%%%%%%%%%%%%%%%%%%%%%%%%%%%%%%%%%%%%%%%%%%%%%%%%%%
% %%%%%%%%%%%%%%%%%%%%%%%%%%%%%%%%%%%%%%%%%%%%%%%%%%%%%%%%%%%%%%%%%%%%%%%%%%%%%%%%%%%%%%%%%%
\section{The Model}
\label{sec:model}
% %%%%%%%%%%%%%%%%%%%%%%%%%%%%%%%%%%%%%%%%%%%%%%%%%%%%%%%%%%%%%%%%%%%%%%%%%%%%%%%%%%%%%%%%%%
% %%%%%%%%%%%%%%%%%%%%%%%%%%%%%%%%%%%%%%%%%%%%%%%%%%%%%%%%%%%%%%%%%%%%%%%%%%%%%%%%%%%%%%%%%%
Our model is obtained by reducing the ${\cal N}=1$ pure super SU($N$)
Yang-Mills theory in $D=6$ dimensions to a point. One obtains a
matrix model with $D=6$ bosonic traceless Hermitian $ N\times N$
matrices $(A_\mu)_{ij}$, $\mu=1,\ldots,6$, $i,j=1,\ldots,N$ and
$2^{D/2-1}=4$ fermionic traceless $ N\times N$ matrices with Grassmann
entries $(\psi_\alpha)_{ij}$, $\alpha=1,\ldots,4$. The action
$\Sb+\Sf$ is given by the bosonic part \Sb and the fermionic part
\Sf
%% %%%%%%%%%%%%%%%
\begin{eqnarray}
\Sb &=& - \frac{1}{4} N\tr [A_\mu,A_\nu]^2\label{m01a}\\
\Sf &=& N
    \tr\left(
    \bar\psi_\alpha(\Gamma_\mu)_{\alpha\beta}[A_\mu,\psi_\beta]
    \right)\label{m01b}\, ,
\end{eqnarray}
%% %%%%%%%%%%%%%%%
and the model is defined by the partition function
%% %%%%%%%%%%%%%%%
\begin{equation}
\label{m02}
Z = \int dA d\psi d\bar\psi\,{\rm e}^{-(\Sb+\Sf)}\, .
\end{equation}
%% %%%%%%%%%%%%%%%
The action is invariant under SO(6) rotations, under which $A_\mu$
transform as a vector and $\psi_\alpha$ as a Weyl spinor. The 
$4\times 4$ gamma matrices $\Gamma_\mu$ ontained after Weyl projection
can be taken to be
%% %%%%%%%%%%%%%%%
\begin{eqnarray}
\Gamma_1 &=& i\sigma_1  \otimes\sigma_2  \, ,\quad
\Gamma_2  =  i\sigma_2  \otimes\sigma_2  \, ,\quad
\Gamma_3  =  i\sigma_3  \otimes\sigma_2  \, ,\nn\\
\Gamma_4 &=& i\mathbf{1}\otimes\sigma_1  \, ,\quad
\Gamma_5  =  i\mathbf{1}\otimes\sigma_3  \, ,\quad
\Gamma_6  =   \mathbf{1}\otimes\mathbf{1}\, ,
\label{m03}
\end{eqnarray}
%% %%%%%%%%%%%%%%%
where $\sigma_i$, $i=1,2,3$ are the Pauli matrices.

Integrating out the fermionic degrees of freedom, we obtain
%% %%%%%%%%%%%%%%%
\begin{equation}
\label{m04}
\detM = \int d\psi d\bar\psi\, {\rm e}^{-S_f}\, ,
\end{equation}
%% %%%%%%%%%%%%%%%
where $\cal M$ is a $4(N^2-1)\times 4(N^2-1)$ matrix, representing the
linear transformation
%% %%%%%%%%%%%%%%%
\begin{equation}
\label{m05}
\Psi_\alpha \mapsto ({\cal M}\Psi)_\alpha 
\equiv 
(\Gamma_\mu)_{\alpha\beta} [A_\mu,\Psi_\beta]\, ,
\end{equation}
%% %%%%%%%%%%%%%%%
acting on the linear space of traceless complex $N\times N$ matrices
$\Psi_\alpha$. The determinant $\detM$ takes complex values in general
and we define its phase $\Gamma$ by $\detM = |\detM|\, {\rm
  e}^{i\Gamma}$.

Eq. \rf{m02} becomes
%% %%%%%%%%%%%%%%%
\begin{equation}
\label{m06}
Z = \int dA\, {\rm e}^{-\Sb}\detM 
  = \int dA\, {\rm e}^{-S}\, ,
\end{equation}
%% %%%%%%%%%%%%%%%
where the effective action is
%% %%%%%%%%%%%%%%%
\begin{equation}
\label{m07}
S = \Sb - \ln\detM\, .
\end{equation}
%% %%%%%%%%%%%%%%%

In \cite{Nishimura:2000ds}, it was shown that $A_\mu$ configurations that are
close to  $d$-dimensional
configurations $3\leq d\leq 6$ leads to milder fluctuations of
$\Gamma$ for smaller values of $d$. A ``$d$--dimensional
configuration'' is one that, by an appropriate SO(6) transformation, we
can set $A_{d+1}=\ldots = A_6 = 0$. For $d=2$, we have that $\detM=0$,
showing that these configurations are suppressed in Eq. \rf{m06}. This
indicates that SO(6) maybe broken down to SO(3), but whether this
is realized is a dynamical question depending on the competition with the
larger entropy of configurations close to higher dimensional
configurations. This question was addressed using the GEM in
\cite{Aoyama:2010ry}, where the free energy of the SO($d$) vacuum was
calculated up to fifth order and it was found that the SO(3) vacuum has
the lowest free energy, which implies SSB to SO(3). The extent of spacetime
%% %%%%%%%%%%%%%%%
\begin{equation}
\label{m08}
\lambda_\mu = \frac{1}{N}\,\tr (A_\mu)^2\, 
\end{equation}
%% %%%%%%%%%%%%%%%
in the SO($d$) vacuum has expectation values $\vev{\lambda_\mu}$ which
are large in $d$ directions and small in the remaining $(6-d)$
directions. This was calculated up to fifth order in the GEM and the
result for the SO(3) vacuum is
%% %%%%%%%%%%%%%%%
\begin{equation}
\vev{\lambda_\mu} \approx \left\{
\begin{array}{l l}
  1.7 & \mbox{for the three extended  directions,}\\
  0.2 & \mbox{for the three shrunken directions.}
\end{array}
\right.
\label{m09}
\end{equation}
%% %%%%%%%%%%%%%%%

% %%%%%%%%%%%%%%%%%%%%%%%%%%%%%%%%%%%%%%%%%%%%%%%%%%%%%%%%%%%%%%%%%%%%%%%%%%%%%%%%%%%%%%%%%%
% %%%%%%%%%%%%%%%%%%%%%%%%%%%%%%%%%%%%%%%%%%%%%%%%%%%%%%%%%%%%%%%%%%%%%%%%%%%%%%%%%%%%%%%%%%
\section{The CLM Applied to the 6D Type IIB Matrix Model}
\label{sec:clm}
% %%%%%%%%%%%%%%%%%%%%%%%%%%%%%%%%%%%%%%%%%%%%%%%%%%%%%%%%%%%%%%%%%%%%%%%%%%%%%%%%%%%%%%%%%%
% %%%%%%%%%%%%%%%%%%%%%%%%%%%%%%%%%%%%%%%%%%%%%%%%%%%%%%%%%%%%%%%%%%%%%%%%%%%%%%%%%%%%%%%%%%
The degrees of freedom in the model given by Eq. \rf{m06} are the
Hermitian traceless matrices $A_\mu$. The complex Langevin equation is the
stochastic differential equation in the fictitious time $t$ involving
the {\it general complex} traceless matrices $A_\mu(t)$
\cite{Parisi:1984cs,Klauder:1983sp}
%% %%%%%%%%%%%%%%%
\begin{equation}
\label{c01}
\frac{d(A_\mu)_{ij}(t)}{dt} =
-\left.\frac{\partial S}{\partial(A_\mu)_{ji}}
 \right|_{A_\mu=A_\mu(t)}
+(\eta_\mu)_{ij}(t)\, ,
\end{equation}
%% %%%%%%%%%%%%%%%
where the $\eta_\mu(t)$ are traceless Hermitian matrices whose elements
are random variables obeying the Gaussian distribution 
$\propto\exp\left(-\dfrac{1}{4}\int\tr\{\eta_\mu(t)\}^2\, dt\right)$
and $S$ is the effective action \rf{m07}.
The drift term in the above equation is the term
$-{\partial S}/{\partial(A_\mu)_{ji}}$,
which is given explicitly by
%% %%%%%%%%%%%%%%%
\begin{equation}
\label{c02}
\frac{\partial   S}{\partial(A_\mu)_{ji}} =
\frac{\partial \Sb}{\partial(A_\mu)_{ji}}
-\Tr\left(
\frac{\partial {\cal M}}{\partial(A_\mu)_{ji}}{\cal M}^{-1}
\right)\, ,
\end{equation}
%% %%%%%%%%%%%%%%%
where $\Tr$ represents the trace of a $4(N^2-1)\times 4(N^2-1)$
matrix. The second term in the above equation is not Hermitian, which
makes the use of general complex traceless matrices $A_\mu(t)$
necessary. The expectation value $\vev{{\cal O}[A_\mu]}$ of an
observable ${\cal O}[A_\mu]$, 
%% %%%%%%%%%%%%%%%
\begin{equation}
\label{c03}
\vev{{\cal O}[A_\mu]}=\frac{1}{Z}
 \int dA\,{\rm e}^{-S}\, ,
\end{equation}
%% %%%%%%%%%%%%%%%
can be calculated from a solution of Eq. \rf{c02} from
%% %%%%%%%%%%%%%%%
\begin{equation}
\label{c04}
\vev{{\cal O}[A_\mu]}=
\frac{1}{T}\int_{t_0}^{t_0+T} {\cal O}[A_\mu(t)]dt\, ,
\end{equation}
%% %%%%%%%%%%%%%%%
where $t_0$ is the thermalization time and $T$ is large enough in
order to obtain satisfactory statistics.

In order for the Eq. \rf{c03} and Eq. \rf{c04} to give the same
result, the  probability distribution $P(A^{(R)}_\mu,A^{(I)}_\mu;t)$ of the
(general complex traceless matrix) solutions $A_\mu(t)$ of Eq. \rf{c01}, where
$A_\mu^{(R)}(t)=(A_\mu(t)+A_\mu^\dagger(t))/2$, $A_\mu^{(I)}(t)=(A_\mu(t)-A_\mu^\dagger(t))/2i$, 
 must satisfy the relation 
%% %%%%%%%%%%%%%%%
\begin{equation}
\label{c05}
\int dA_\mu\, \rho(A_\mu;t) {\cal O}[A_\mu] =
\int dA_\mu^{(R)}dA_\mu^{(I)}\, P(A_\mu^{(R)},A_\mu^{(I)};t) 
     {\cal O}[A_\mu^{(R)}+i A_\mu^{(I)}]\, .
\end{equation}
%% %%%%%%%%%%%%%%%
On the LHS of the above equation, $A_\mu$ are the original Hermitian
matrices in the model given by Eq. \rf{m06} and $\rho(A_\mu;t)$ is a complex
weight which is a solution of a Fokker-Planck equation, such that
$\lim\limits_{t\to\infty}\rho(A_\mu;t) = {\rm e}^{-S}/Z$, giving the desired
$\vev{{\cal O}[A_\mu]}$ in the $t\to\infty$ limit. On the RHS of
Eq. \rf{c05}, we have the (real positive)
probability distribution 
$P(A_\mu^{(R)},A_\mu^{(I)};t)$
of the complex matrix solutions $A_\mu(t)$ of Eq. \rf{c01} and the analytic continuation of
${\cal O}[A_\mu] \mapsto {\cal O}[A_\mu^{(R)}+i A_\mu^{(I)}]$. For large enough $t$, the RHS of
Eq. \rf{c05} is calculated using the RHS of Eq. \rf{c04}. A
necessary and sufficient condition for the equality in Eq. \rf{c04} to hold is that
the probability distribution of 
%% %%%%%%%%%%%%%%%
\begin{equation}
\label{c06}
u = \sqrt{\frac{1}{6 N^3}\sum_{\mu=1}^{6} \sum_{i,j=1}^N
\left| \frac{\partial S}{\partial (A_\mu)_{ij}}  \right|^2 }\, ,
\end{equation}
%% %%%%%%%%%%%%%%%
in $P(A^{(R)},A^{(I)};t)$ falls off exponentially or faster\cite{Nagata:2016vkn}.
There are two basic reasons for violating this condition. The first
one is the ``excursion problem'', where the solutions of Eq. \rf{c01}
drift deep into the anti--Hermitian direction. The second one is the
``singular drift problem'', which occurs due to the appearance of
${\cal M}^{-1}$ in Eq. \rf{c02} when some eigenvalues of $\cal M$
accumulate near zero frequently. 

The excursion problem can be avoided by using the gauge cooling
technique\cite{Seiler:2012wz}. We minimize the ``Hermitian norm''
%% %%%%%%%%%%%%%%%
\begin{equation}
\label{c07}
{\cal N}_H(t) = -\frac{1}{6 N}
\sum_{\mu=1}^6 \tr\left\{
\left( A_\mu(t) - A_\mu(t)^\dagger\right)^2
\right\}\, ,
\end{equation}
%% %%%%%%%%%%%%%%%
by performing an SL($N,\mathds{C}$) transformation 
$A_\mu(t)\mapsto g(t) A_\mu g(t)^{-1}$, where $g(t)=\exp\{-\alpha G(t)\}$ and
$G(t)=\dfrac{1}{N}\sum\limits_{\mu=1}^6 [A_\mu(t),A_\mu(t)^\dagger]$. 
$G(t)$ is the gradient of ${\cal N}_H(t)$ wrt the  SL($N,\mathds{C}$)
transformation\cite{Ito:2016efb}. The real positive parameter $\alpha$
is computed so that    ${\cal N}_H(t)$ is minimized. In
\cite{Nagata:2016vkn,Nagata:2015uga}, it was shown that gauge cooling
does not affect the argument for the justification of the CLM.

\begin{figure}[htbp]
\centering 
\includegraphics[width=0.48\textwidth]{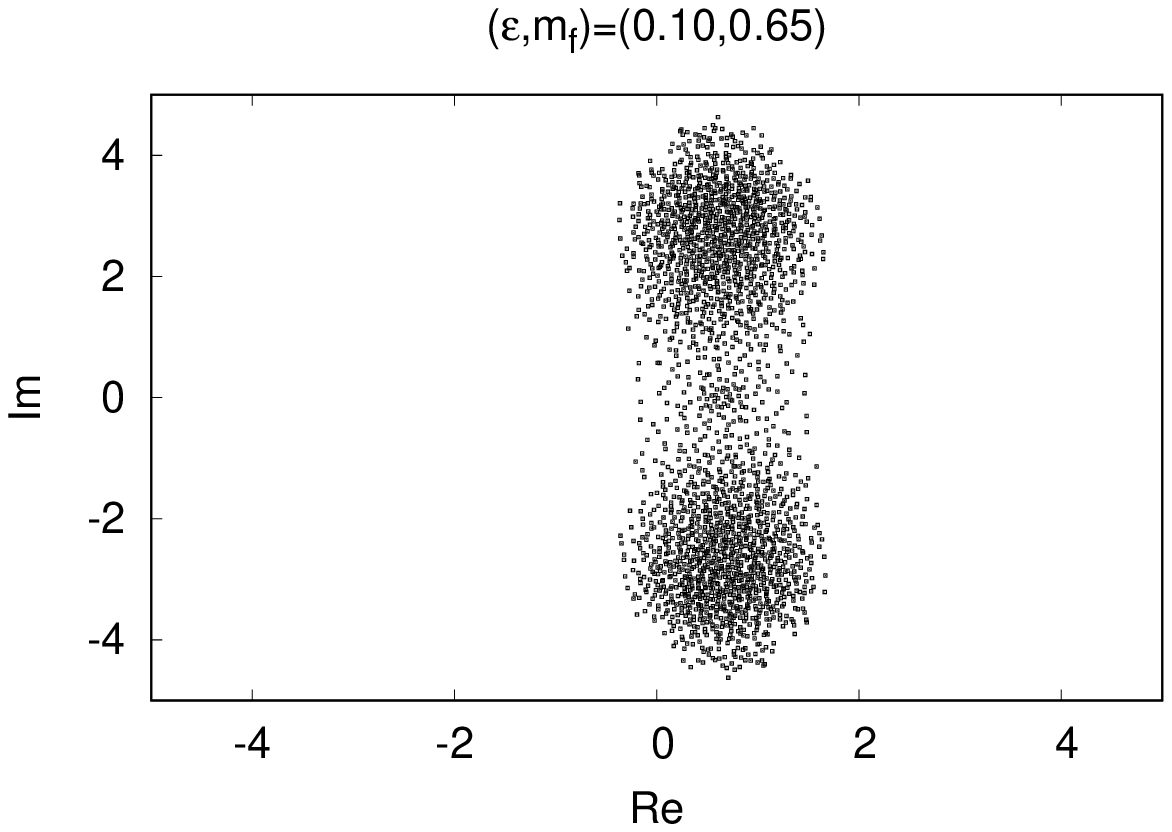}
\includegraphics[width=0.48\textwidth]{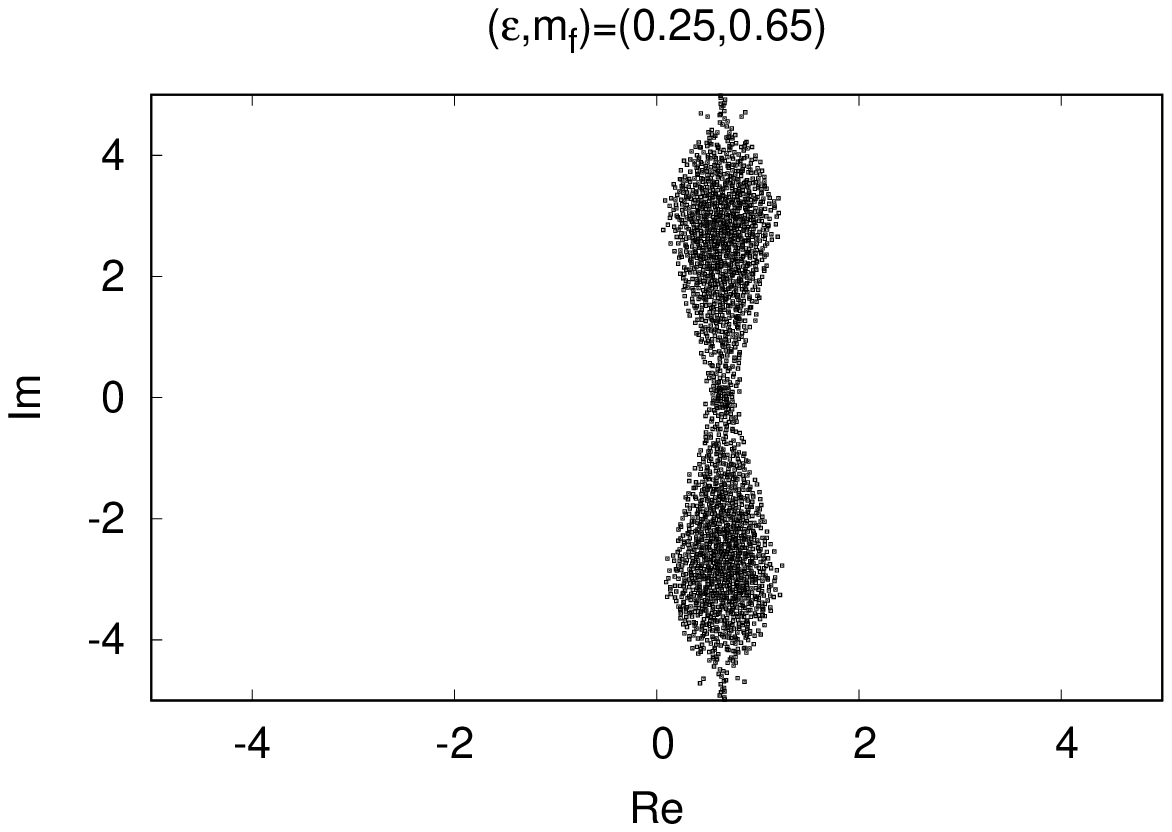}
     \caption{The effect of the deformation $\Delta\Sf$ of Eq. \protect\rf{c08} on
       the eigenvalues of the matrix $\cal M$ for a typical
       configuration for $N=24$, $\mf=0.65$, $\epsilonm=0.10$ (Left)
       and $\epsilonm=0.25$ (Right). By increasing $\mf$, the
       eigenvalues shift in the direction of the real axis. Notice also
       that by increasing $\epsilonm$ for given $\mf$, the spread of
       the eigenvalues in the real direction decreases.\label{f:1}}
\end{figure}

The singular drift problem can be avoided by deforming the fermionic
action by adding the term
%% %%%%%%%%%%%%%%%
\begin{equation}
\label{c08}
\Delta\Sf = N\mf\tr\left(
\bar\psi_\alpha (\Gamma_6)_{\alpha\beta} \psi_\beta
\right)\, 
\end{equation}
%% %%%%%%%%%%%%%%%
to the action, so that $\Sf\mapsto \Sf+\Delta\Sf$.  $\mf\geq 0$
is the deformation parameter. This term modifies the matrix
$\cal M$ of Eq. \rf{m05}, so that
%% %%%%%%%%%%%%%%%
\begin{equation}
\label{c08a}
\Psi_\alpha \mapsto ({\cal M}\Psi)_\alpha 
\equiv 
(\Gamma_\mu)_{\alpha\beta} [A_\mu,\Psi_\beta]+\mf \Psi_\alpha\, ,
\end{equation}
%% %%%%%%%%%%%%%%%
and shifts its eigenvalues in the real direction. A typical case is shown in figure \ref{f:1}.
This method was successfully applied
in \cite{Ito:2016efb} in an SO(4) symmetric matrix model with a
complex fermion determinant and a severe complex action problem. 
% %%%%%%%%%%%%%%%%%%%%%%%% Figure %%%%%%%%%%%%%%%%%%%%%%%%%%%%%%%%%%%%%
% %%%%%%%%%%%%%%%%%%%%%%%%%%%%%%%%%%%%%%%%%%%%%%%%%%%%%%%%%%%%%%%%%%%%%
For
$\mf$ large enough, the eigenvalues of $\cal M$ avoid zero and we
don't have the singular drift problem. This term breaks the SO(6)
symmetry down to SO(5). Since we are looking for SO($d$) SSB patterns
with $d<5$, this is not a problem. In the end, the $\mf\to 0$ limit
will be taken in order to obtain the $D=6$ IIB model. Assuming that
nothing dramatic happens in the $\mf=0$ region, the extrapolation from
the region of small $\mf$ will give the correct pattern of SSB of the
undeformed model, as it happened in \cite{Ito:2016efb}, where the
expected SO(4) to SO(2) breaking was observed. When
$\mf\to\infty$, the fermions decouple and we obtain a matrix model
with only bosonic degrees of freedom. In this case, it is known that
the SO(6) symmetry is not broken \cite{Hotta:1998en}. Therefore, the
deformation parameter \mf can be seen as interpolating between the
bosonic matrix model and the  $D=6$ IIB model.

The numerical solution of Eq. \rf{c01} is computed by discretizing the
fictitious time $t$
%% %%%%%%%%%%%%%%%
\begin{equation}
\label{c09}
(A_\mu)_{ij}(t+\Delta t) = 
(A_\mu)_{ij}(t)
-\Delta t \left.\frac{\partial S}{\partial(A_\mu)_{ji}}
 \right|_{A_\mu=A_\mu(t)}
+\sqrt{\Delta t}\,(\eta_\mu)_{ij}(t)\, .
\end{equation}
%% %%%%%%%%%%%%%%%
The $\sqrt{\Delta t}$ comes from the chosen normalization of the
$\eta_\mu(t)$ 
$\propto \exp\left(-\dfrac{1}{4}\sum\limits_j\tr\{\eta_\mu(t)\}^2\right)$.
The stepsize $\Delta t$ is chosen adaptively, so that the drift term
remains small\cite{Aarts:2009dg}. The details of the numerical
computation can be found in \cite{Anagnostopoulos:2017gos}. 

The order parameters of the SSB are taken to be the spacetime
extensions in the $\mu$--direction
%% %%%%%%%%%%%%%%%
\begin{equation}
\label{c10}
\lambda_\mu = \frac{1}{N} \, \tr\left(A_\mu\right)^2\, ,
\end{equation}
%% %%%%%%%%%%%%%%%
where no sum over $\mu$ is taken. In order to calculate
$\vev{\lambda_\mu}$, we add the term
%% %%%%%%%%%%%%%%%
\begin{equation}
\label{c11}
\Delta \Sb = \frac{1}{2} N \epsilonm
\sum_{\mu=1}^6 m_\mu \tr\left(A_\mu\right)^2
\end{equation}
%% %%%%%%%%%%%%%%% 
to the action, so that $\Sb\mapsto\Sb+\Delta\Sb$, where we take 
$0<m_1\leq\ldots\leq m_6$ and $\epsilonm>0$. This term breaks the SO(6) symmetry
explicitly, and SSB is probed by first taking the large--$N$ limit and
then sending $\epsilonm\to 0$. Notice that, although $\lambda_\mu(t)$
is not real for a configuration $A_\mu(t)$, the expectation values
$\vev{\lambda_\mu}$ {\it are} real due to the symmetry of the drift
term \rf{c02} under $A_i\mapsto A_i^\dagger$ for $i=1, \ldots, 5$ and
$A_6\mapsto -A_6^\dagger$. Due to the chosen ordering of the $m_\mu$,
we will have that 
%% %%%%%%%%%%%%%%%
\begin{equation}
\label{c12}
\vev{\lambda_1}\geq\vev{\lambda_2}\geq\ldots\geq\vev{\lambda_6}\, .
\end{equation}
%% %%%%%%%%%%%%%%%
When we take the large-$N$ limit and then $\epsilonm\to 0$ and find
that the $\vev{\lambda_\mu}$ are not equal, we conclude that the SO(6)
symmetry is spontaneously broken. For finite $\mf$, SSB occurs
if we find that some of the $\vev{\lambda_\mu}$ are not equal for
$\mu=1,\ldots,5$. 

In this work, we take
%% %%%%%%%%%%%%%%%
\begin{equation}
\label{c13}
m_\mu = (0.5, 0.5, 1, 2, 4, 8)\, .
\end{equation}
%% %%%%%%%%%%%%%%%
This choice retains the SO(2) symmetry, but since we do not expect the
SSB to SO(2) to occur, this is not a problem. It is preferable to keep
the spectrum of the $m_\mu$ not too wide in order to take the
$\epsilonm\to 0$ extrapolation without introducing large systematic
errors.

% %%%%%%%%%%%%%%%%%%%%%%%%%%%%%%%%%%%%%%%%%%%%%%%%%%%%%%%%%%%%%%%%%%%%%%%%%%%%%%%%%%%%%%%%%%
% %%%%%%%%%%%%%%%%%%%%%%%%%%%%%%%%%%%%%%%%%%%%%%%%%%%%%%%%%%%%%%%%%%%%%%%%%%%%%%%%%%%%%%%%%%
\section{Results}
\label{sec:res}
% %%%%%%%%%%%%%%%%%%%%%%%%%%%%%%%%%%%%%%%%%%%%%%%%%%%%%%%%%%%%%%%%%%%%%%%%%%%%%%%%%%%%%%%%%%
% %%%%%%%%%%%%%%%%%%%%%%%%%%%%%%%%%%%%%%%%%%%%%%%%%%%%%%%%%%%%%%%%%%%%%%%%%%%%%%%%%%%%%%%%%%
To summarize, the model that we investigate numerically using the CLM is
given by Eq. \rf{c09}, where we have taken 
$S\mapsto S+\Delta\Sb+\Delta\Sf$. We use \rf{c04} to compute the
expectation values $\vev{\lambda_\mu}_{\mf,\epsilonm,N}$. In order to check for the large
excursion and singular drift problems, we measure the norm 
${\cal N}_H$ of Eq. \rf{c07} and the magnitude of the drift $u$ of
Eq. \rf{c06} and plot their histograms and time histories. 
% %%%%%%%%%%%%%%%%%%%%%%%% Figure %%%%%%%%%%%%%%%%%%%%%%%%%%%%%%%%%%%%%
\begin{figure}[htbp]
\centering 
\includegraphics[width=0.48\textwidth]{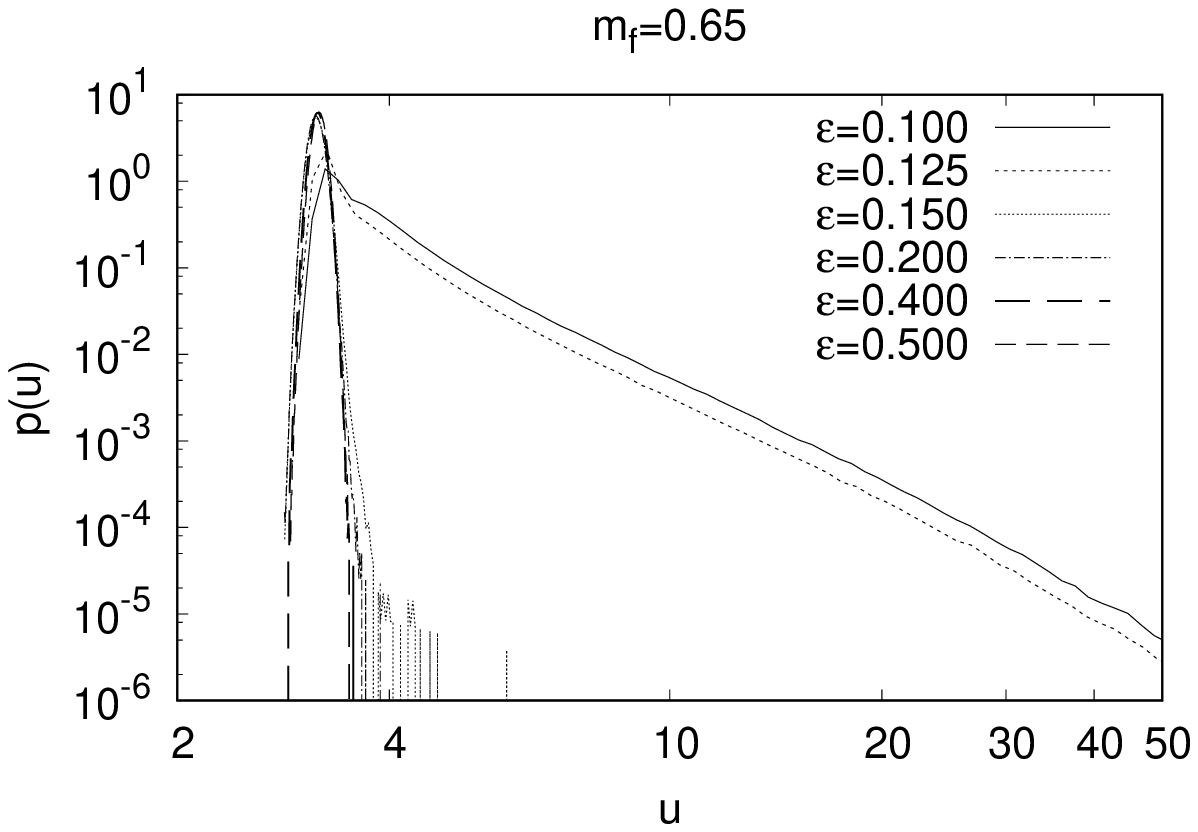}
\includegraphics[width=0.48\textwidth]{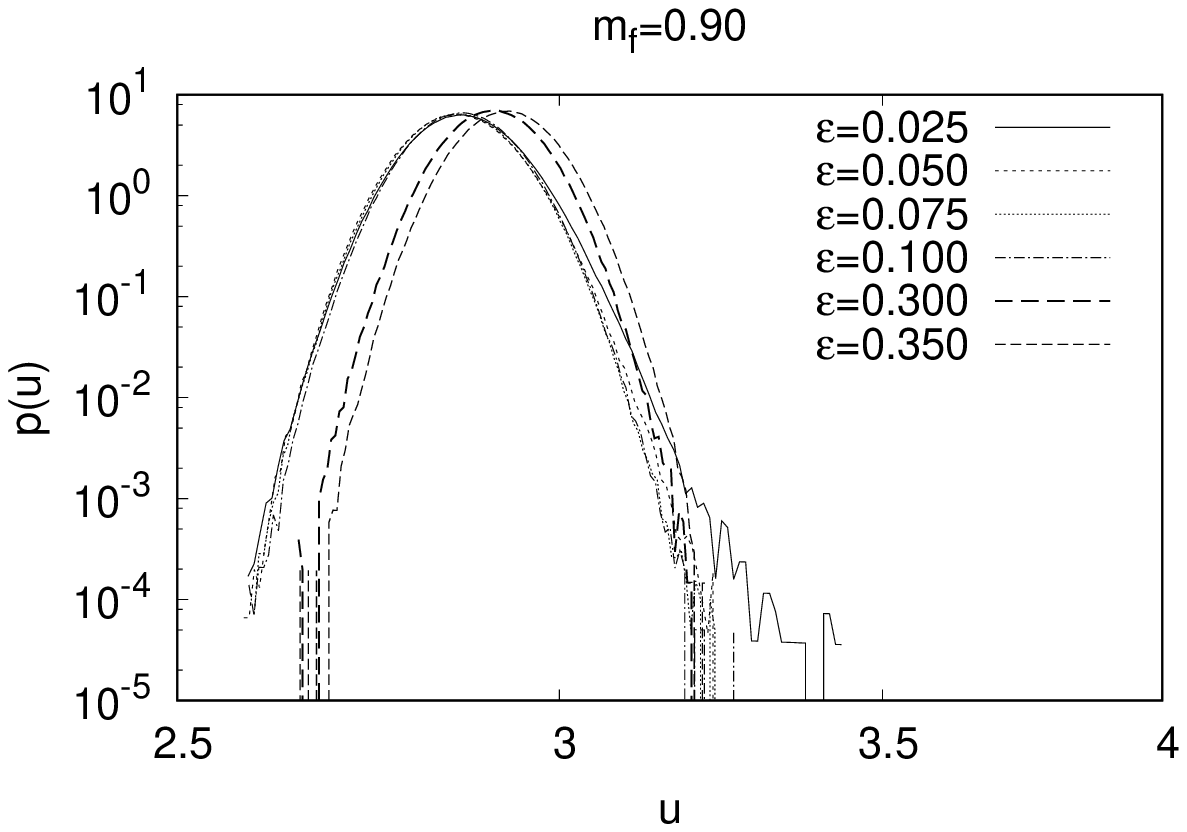} % \mf = 0.90
  \caption{The probability distribution $p(u)$ of $u$ defined in
    Eq. \protect\rf{c06} for $N=24$ with $\mf=0.65$ (Left) and
    $\mf=0.90$ (Right). \label{f:2}}
\end{figure}
% %%%%%%%%%%%%%%%%%%%%%%%%%%%%%%%%%%%%%%%%%%%%%%%%%%%%%%%%%%%%%%%%%%%%%
In figure \ref{f:2} we plot the histogram $p(u)$ for $N=24$ and
$\mf=0.65, 0.90$. For $\mf=0.65$, we see that $p(u)$ falls off
exponentially or faster for $\epsilonm\geq 0.150$, whereas it develops
a power-law tail for $\epsilonm \leq 0.125$. Therefore, we can trust
only the results for  $\epsilonm\geq 0.150$. For $\mf=0.90$ we see
that no power law tail exists for all values of $\epsilonm$ investigated. 

% %%%%%%%%%%%%%%%%%%%%%%%% Figure %%%%%%%%%%%%%%%%%%%%%%%%%%%%%%%%%%%%%
\begin{figure}[htbp]
\centering 
\includegraphics[width=0.48\textwidth]{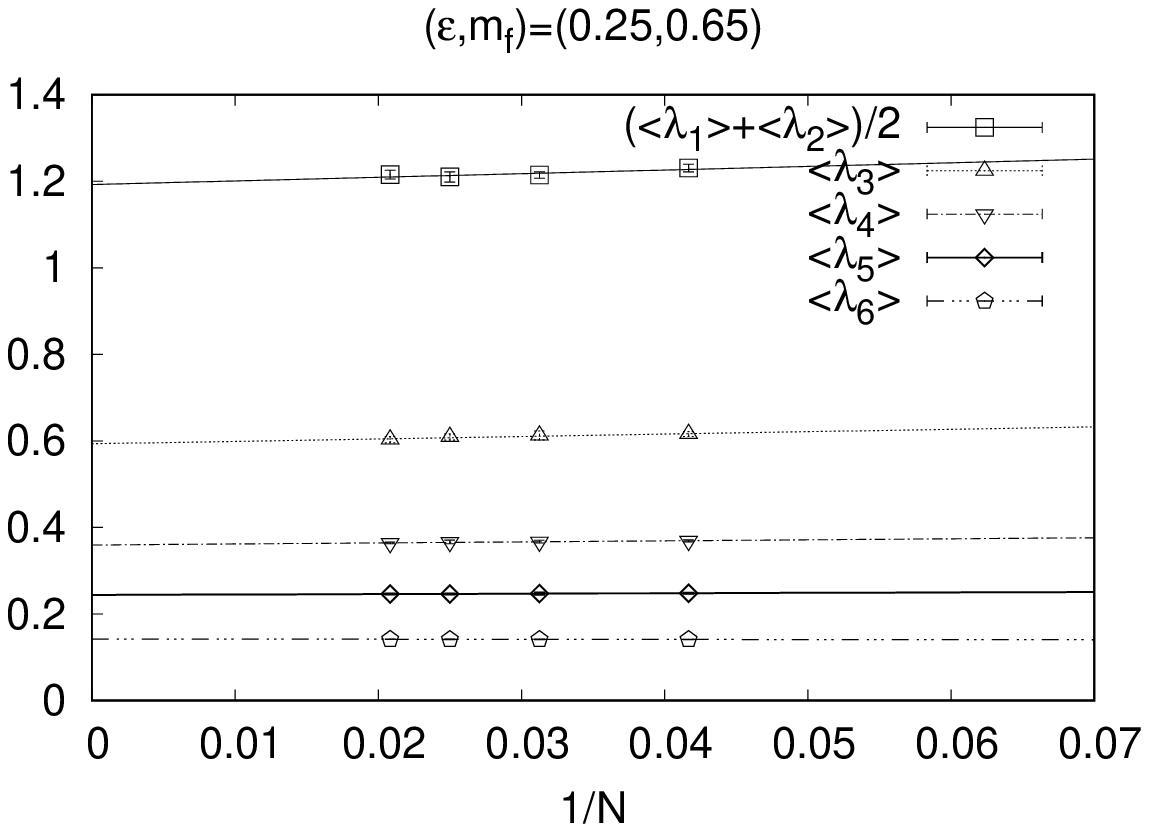} 
  \caption{The expectation values $\vev{\lambda_\mu}_{\mf,\epsilonm,N}$ for $\mf=0.65$,
    $\epsilonm=0.25$ and $N=24, 32, 40, 48$ together with a fit of the
    form $a+b/N$. The fit is used for the large-$N$
    extrapolation discussed in the text and the fitting parameter $a$
    gives  $\vev{\lambda_\mu}_{\mf,\epsilonm}=\lim\limits_{N\to\infty}\vev{\lambda_\mu}_{\mf,\epsilonm,N}$.\label{f:3}}  
\end{figure}
% %%%%%%%%%%%%%%%%%%%%%%%%%%%%%%%%%%%%%%%%%%%%%%%%%%%%%%%%%%%%%%%%%%%%%
In order to probe the SSB, first we have to take the large--$N$ limit
$\vev{\lambda_\mu}_{\mf,\epsilonm}=\lim\limits_{N\to\infty}\vev{\lambda_\mu}_{\mf,\epsilonm,N}$
For that, we plot $\vev{\lambda_\mu}_{\mf,\epsilonm,N}$ as a function
of $1/N$, as in figure \ref{f:3}.  We consider the average
$(\vev{\lambda_1}+\vev{\lambda_2})/2$ instead 
of $\vev{\lambda_1}$ and $\vev{\lambda_2}$ separately due to the choice
\rf{c13} and in order to increase statistics.
The large-$N$ extrapolation is done
by fitting the data to a linear form $a+b/N$. Our data fits nicely for
all values of $(\mf,\epsilonm)$ considered and for $24\leq N \leq 48$
and the coefficient $a$ gives $\vev{\lambda_\mu}_{\mf,\epsilonm}$.

Then we have to take the $\epsilonm\to 0$ limit. We compute the ratio \cite{Ito:2016efb}
%% %%%%%%%%%%%%%%%
\begin{equation}
\label{r01}
\rho_\mu(\mf,\epsilonm)=\frac
{\vev{\lambda_\mu}_{\mf,\epsilonm}}
{\sum\limits_{\nu=1}^6 \vev{\lambda_\nu}_{\mf,\epsilonm}}\, ,
\end{equation}
%% %%%%%%%%%%%%%%%
instead of $\vev{\lambda_\mu}_{\mf,\epsilonm}$, because  some of
the finite $\epsilonm$ effects cancel between the numerator and the
denominator.
% %%%%%%%%%%%%%%%%%%%%%%%% Figure %%%%%%%%%%%%%%%%%%%%%%%%%%%%%%%%%%%%%
\begin{figure}[htbp]
\centering 
\includegraphics[width=0.48\textwidth]{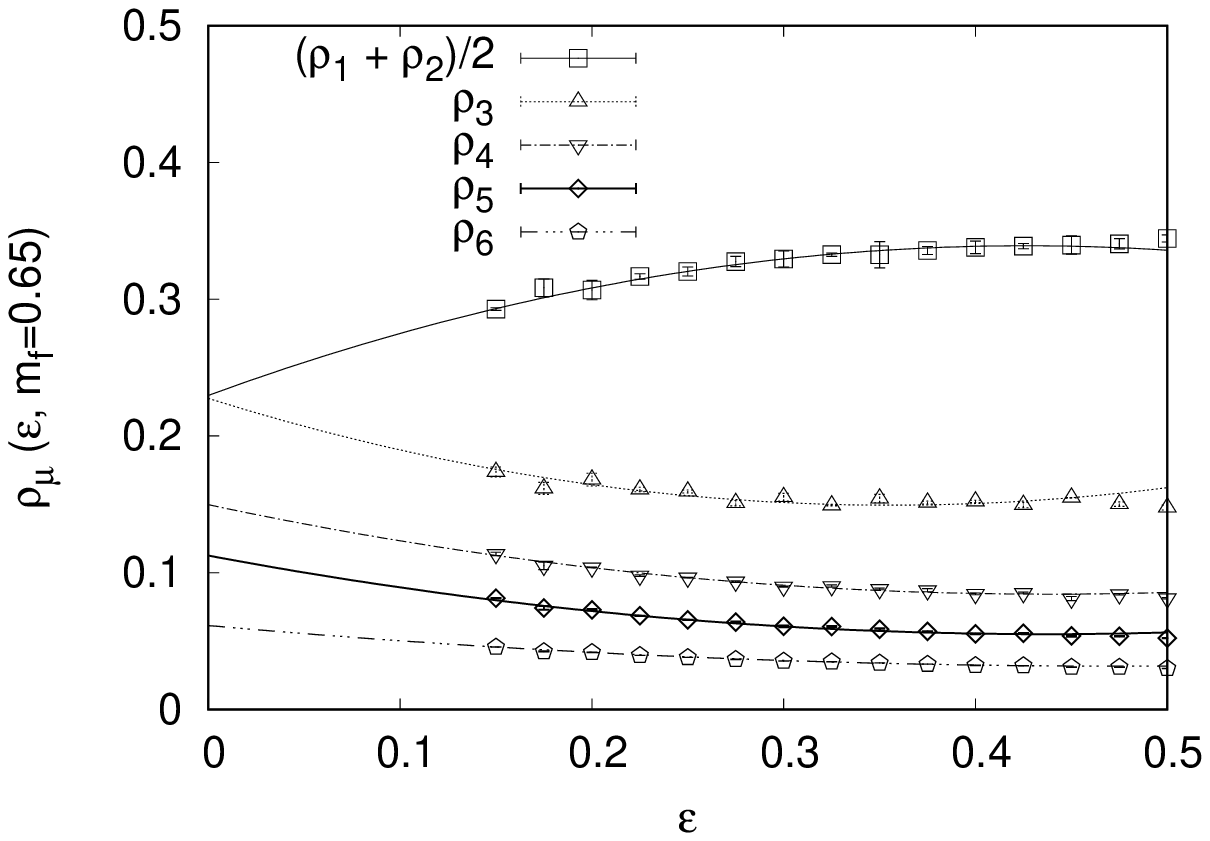}
\includegraphics[width=0.48\textwidth]{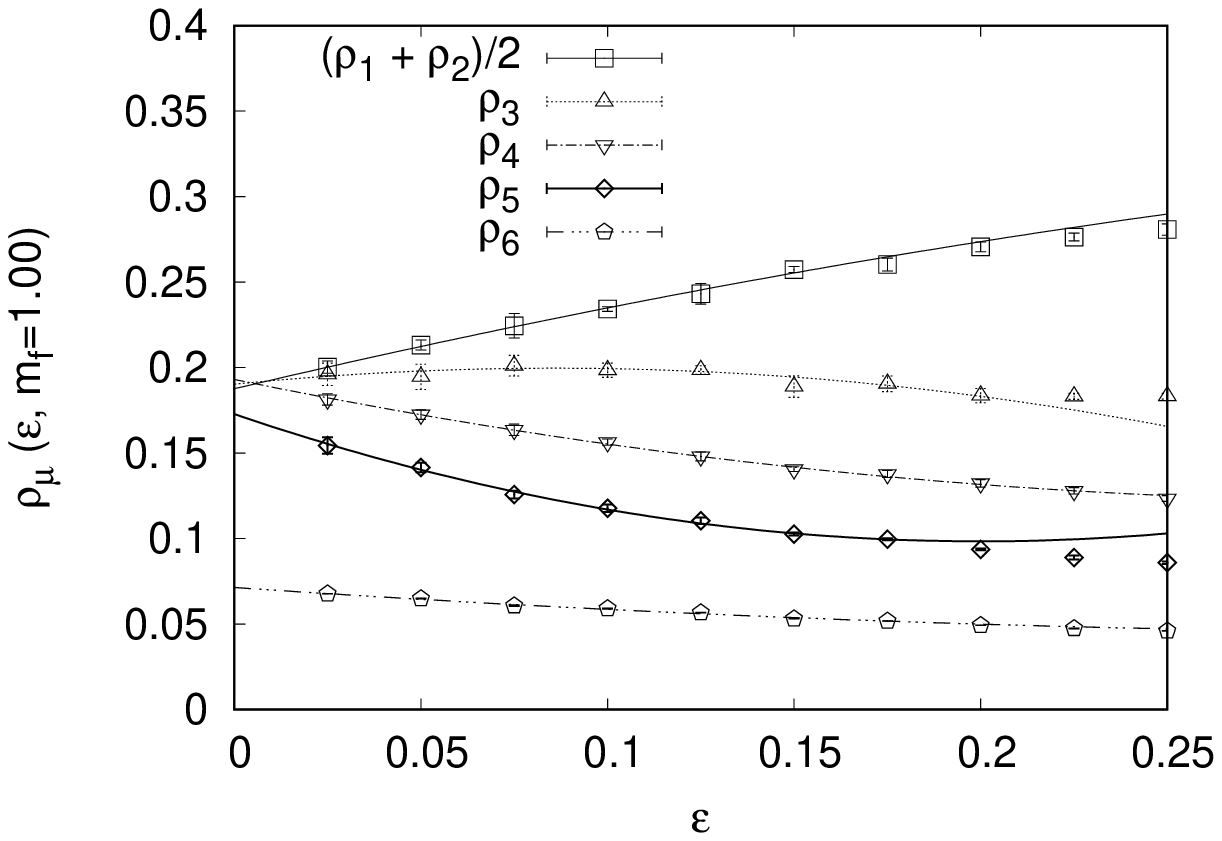}
\includegraphics[width=0.48\textwidth]{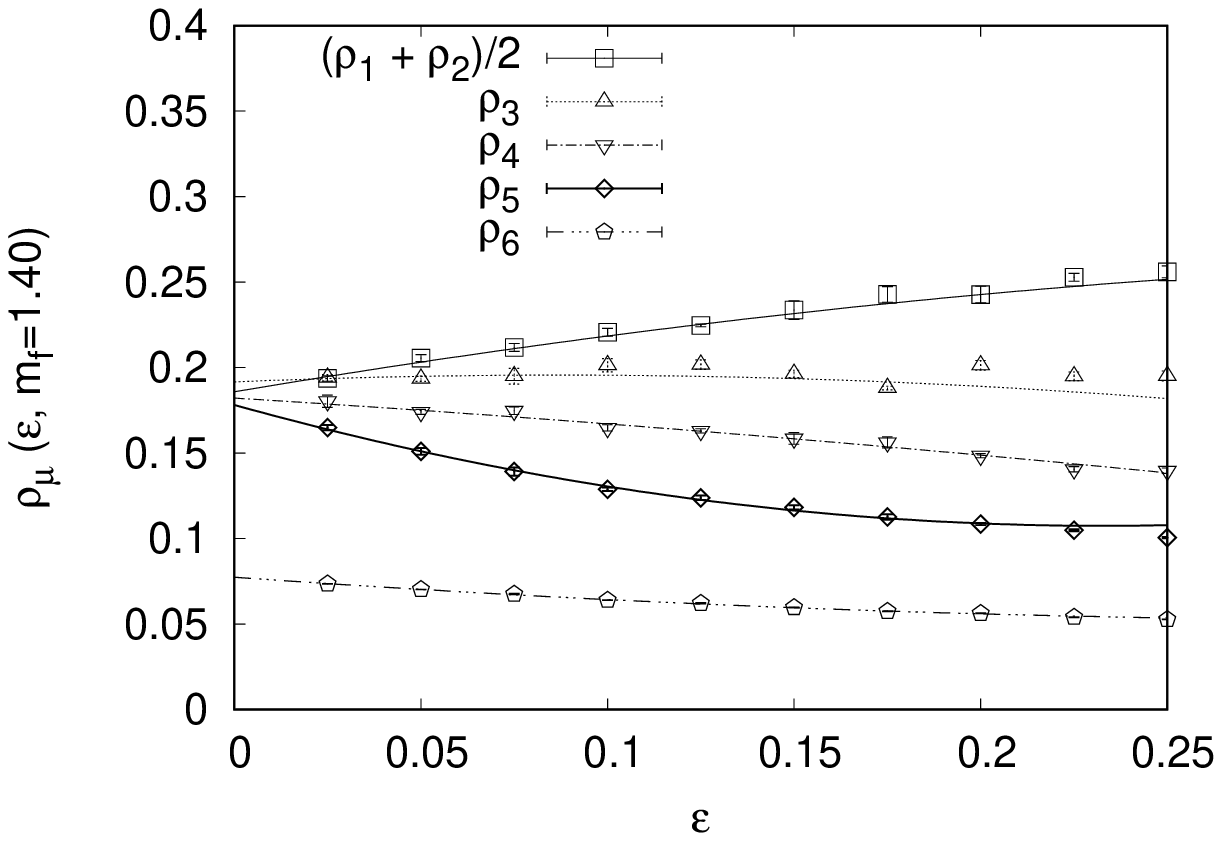}
\includegraphics[width=0.48\textwidth]{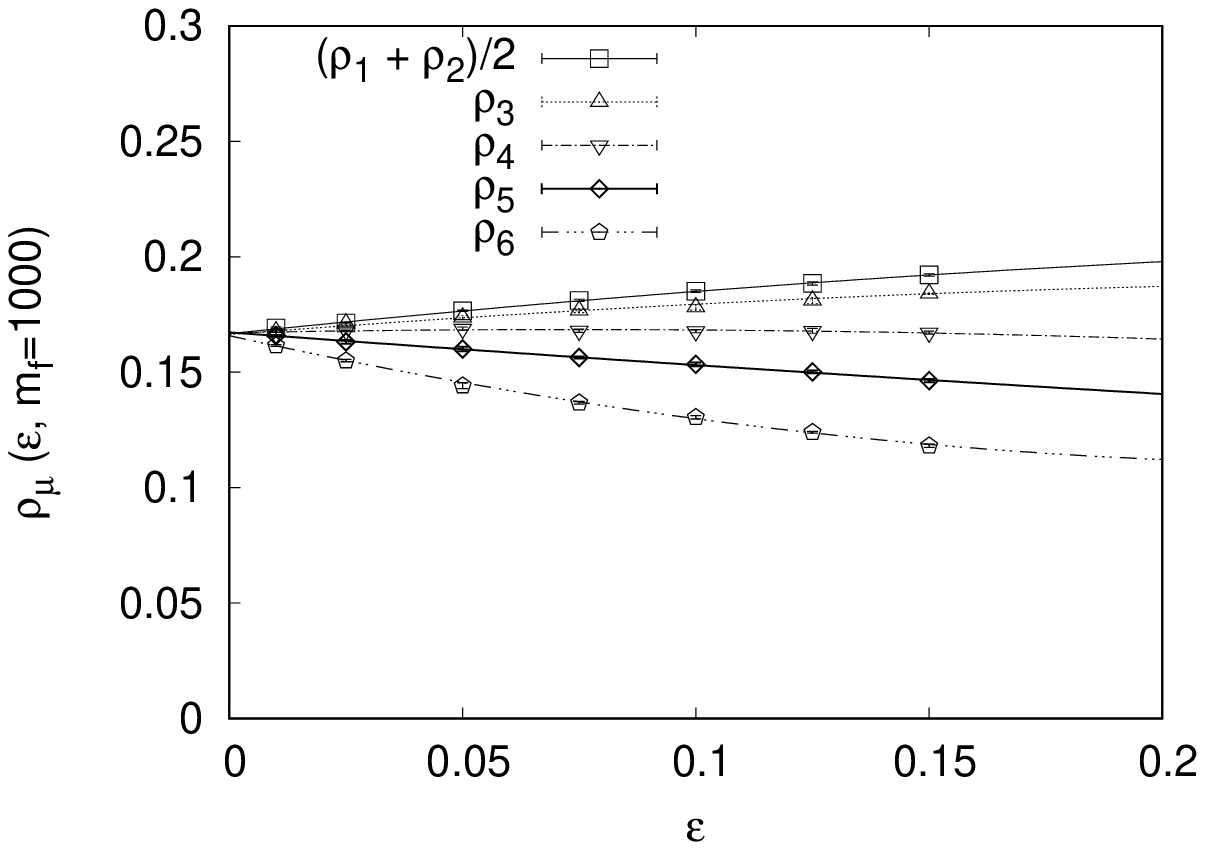}
  \caption{The ratio $\rho_\mu(\mf,\epsilonm)$ defined in
    Eq. \protect\rf{r01} as a function of $\epsilonm$ for $\mf=0.65$
    (Top-Left), $\mf=1.00$ (Top-Right), $\mf=1.40$ (Bottom-Left) and
    $\mf=1000$ (Bottom-Right). Only the values of $\epsilonm$ where
    the singular drift problem is absent are shown. The lines are a
    fit to $a+b\epsilonm + c\epsilonm^2$.\label{f:4}}
\end{figure}
% %%%%%%%%%%%%%%%%%%%%%%%%%%%%%%%%%%%%%%%%%%%%%%%%%%%%%%%%%%%%%%%%%%%%%
In figure \ref{f:4} we show the plots of $\rho_\mu(\mf,\epsilonm)$ as
a function of $\epsilonm$  for $\mf=0.65, 1.00, 1.40, 1000$. We plot
only the data that do not suffer from the singular drift problem by
applying the criterion of exponential or faster suppression of the
tail of $p(u)$ (see figure \ref{f:2}). We also consider the average
$(\rho_1+\rho_2)/2$ instead of $\rho_1$ and $\rho_2$ due to the choice
\rf{c13} and in order to reduce statistical errors. The $\epsilonm\to 0$
extrapolation is done by fitting our data to a quadratic function  
$a+b\epsilonm + c\epsilonm^2$ and the fitting parameter $a$ gives
%% %%%%%%%%%%%%%%%
\begin{equation}
\label{r02}
\rho_\mu(\mf) = \lim_{\epsilonm\to 0}\rho_\mu(\mf,\epsilonm)\, .
\end{equation}
%% %%%%%%%%%%%%%%%

The fitting ranges that satisfy the
singular drift problem criterion and fit well to this function are
$0.150\leq \epsilonm \leq 0.475$ for $\mf=0.65$,
$0.025\leq \epsilonm \leq 0.175$ for $\mf=1.00$,
$0.025\leq \epsilonm \leq 0.200$ for $\mf=1.40$,
$0.010\leq \epsilonm \leq 0.150$ for $\mf=1000$.
For $\mf=0.65$ we see that the curves $(\rho_1+\rho_2)/2$ and $\rho_3$
intersect at $\epsilonm=0$, implying that the SO(5) symmetry of the
deformed model is spontaneously broken to SO(3). 
For $\mf=1.00$ we see that the curves $(\rho_1+\rho_2)/2$, $\rho_3$ and $\rho_4$
intersect at $\epsilonm=0$, implying that the SO(5) symmetry of the
deformed model is spontaneously broken to SO(4). 
For $\mf=1.40$ we see that the curves $(\rho_1+\rho_2)/2$, $\rho_3$, $\rho_4$ and $\rho_5$
intersect at $\epsilonm=0$, implying that the SO(5) symmetry is not
spontaneously broken.
Finally, for $\mf=1000$, all the $\rho_\mu$ curves intersect at
$\epsilonm=0$, implying that the SO(6) symmetry  is not
spontaneously broken. This is consistent with the fact that at
$\mf\to\infty$ the fermions decouple and the deformed model reduces to
the bosonic matrix model.

% %%%%%%%%%%%%%%%%%%%%%%%% Figure %%%%%%%%%%%%%%%%%%%%%%%%%%%%%%%%%%%%%
\begin{figure}[htbp]
\centering 
\includegraphics[width=0.48\textwidth]{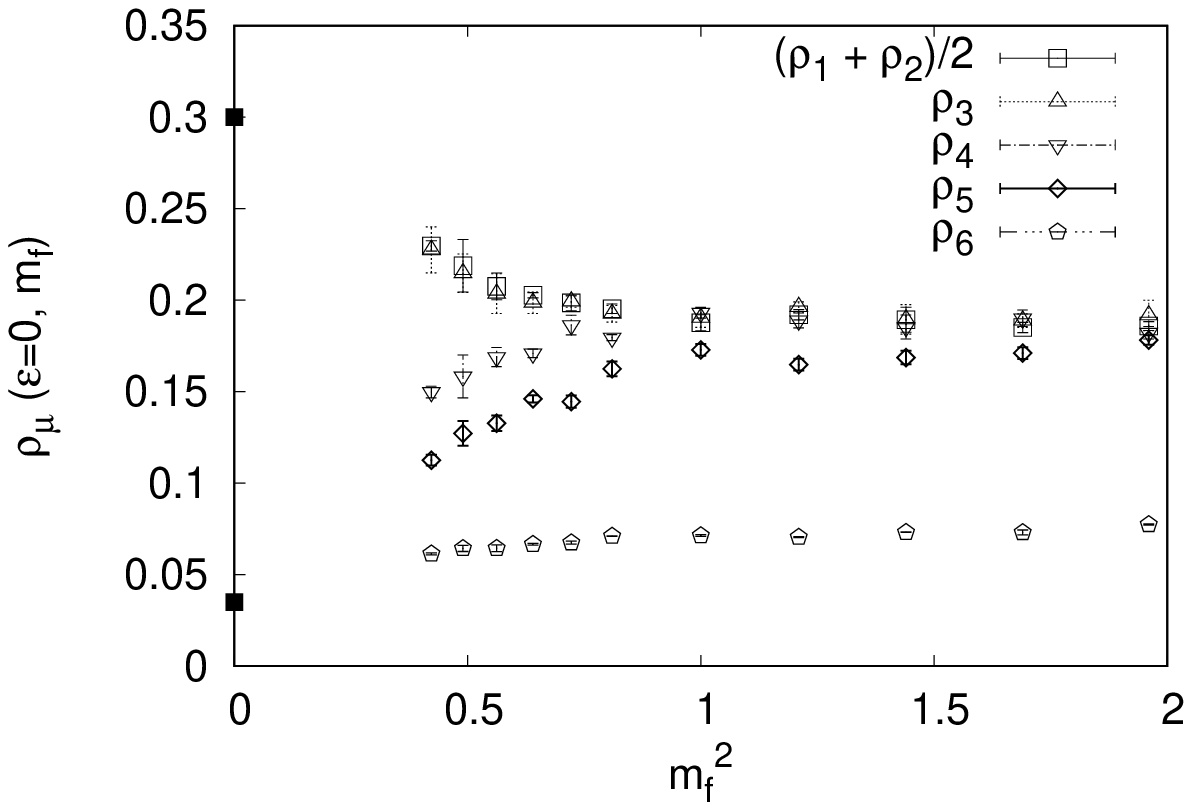}
\includegraphics[width=0.48\textwidth]{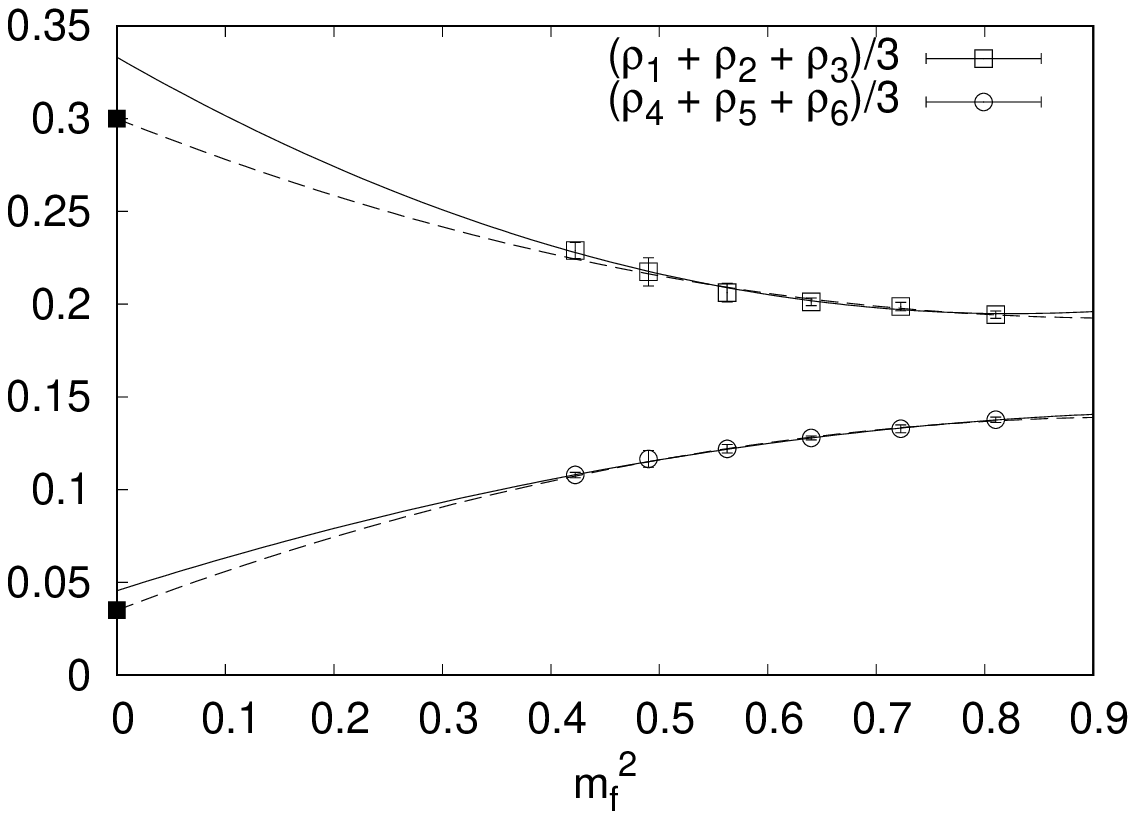}
  \caption{(Left) The values $\rho(\mf)$ of Eq. \protect\rf{r02} as a
    function of $\mf^2$ for $\mf=0.65,0.70, 0.75, 0.80, 0.85, 0.90, 1.0, 1.10, 1.20, 1.30, 1.40$.
    The filled squares are the GEM predictions of Eq. \rf{r03}. 
    (Right) The averages $(\rho_1+\rho_2+\rho_3)/3$ and
    $(\rho_4+\rho_5+\rho_6)/3$ as a
    function of $\mf^2$ for $\mf\leq 0.9$, corresponding to the SO(3)
    symmetric phase. The solid lines are fits to
    $a+b\mf^2+c\mf^4$ and the dashed lines are similar fits
    constrained to pass through the $\mf=0$ points \protect\rf{r03}
    predicted by GEM.\label{f:5}} 
\end{figure}
% %%%%%%%%%%%%%%%%%%%%%%%%%%%%%%%%%%%%%%%%%%%%%%%%%%%%%%%%%%%%%%%%%%%%%
Finally, we consider the $\mf\to 0$ limit, which will give the
undeformed $D=6$ IIB matrix model. In figure \ref{f:5} we plot the
values $\rho_\mu(\mf)$ of Eq. \rf{r02}. We see that an SO(3) vacuum
develops for $\mf\lesssim 0.90$, whereas an SO(4) vacuum develops for
$1.00\lesssim \mf \lesssim 1.30$. Considering the fact that an SO(2)
vacuum does not realize due to the vanishing $\detM$, we conclude that
as $\mf\to 0$ the SO(3) vacuum survives. We conclude that in the
undeformed model $\mf=0$, the SO(6) rotational symmetry is
spontaneously broken to SO(3), in agreement with the GEM prediction.

From Eq. \rf{m09} we obtain
%% %%%%%%%%%%%%%%%
\begin{equation}
\label{r03}
\rho_1=\rho_2=\rho_3\simeq\frac{1.7}{5.7}\simeq 0.3\, ,\quad \quad 
\rho_4\simeq\rho_5\simeq\rho_6\simeq\frac{0.2}{5.7}\simeq 0.035 \ . 
\end{equation}
%% %%%%%%%%%%%%%%%
These values are put in the plots of figure \ref{f:5}. The left plot
of  figure \ref{f:5} shows the averages $(\rho_1+\rho_2+\rho_3)/3$ and
$(\rho_4+\rho_5+\rho_6)/3$ as a function of $\mf^2$ for $\mf\leq 0.9$,
corresponding to the SO(3) symmetric phase. Due to the symmetry
$\mf\mapsto -\mf$, as $\mf\to 0$, the
asymptotic behavior of these functions is expected to be a power
series in $\mf^2$. We fit the corresponding data to a polynomial
$a+b\mf^2+c\mf^4$ for $0.65\leq\mf\leq 0.90$. The $\mf\to 0$
extrapolation gives
%% %%%%%%%%%%%%%%%
\begin{equation}
\label{r04}
\frac{\rho_1+\rho_2+\rho_3}{3}  = 0.33(2) \ , \quad \quad
\frac{\rho_4+\rho_5+\rho_6}{3}  = 0.046(3) \ , 
\end{equation}
%% %%%%%%%%%%%%%%%
which are close to the values \rf{r03} predicted by the GEM. We should
note that the GEM has systematic errors due to the truncations
involved in the calculations. Therefore we conclude that the results
calculated by the CLM Eq. \rf{r04} are in reasonable quantitative
agreement with the GEM results of Eq. \rf{r03}.

\section*{Acknowledgements}
We thank H.~Kawai and H.~Steinacker for valuable comments and discussions. 
This research was supported by MEXT as ``Priority Issue on Post-K computer''
(Elucidation of the Fundamental Laws and Evolution of the Universe) 
and Joint Institute for Computational Fundamental Science (JICFuS). 
Computations were carried out using computational resources such as 
KEKCC, the NTUA het clusters, the FX10 at Kyushu University.
This work was also supported by
computational time granted from the Greek Research \& Technology
Network (GRNET) in the National HPC facility - ARIS - under project ID IKKT10D.
T.~A.\ was supported in part by Grant-in-Aid  for Scientific Research 
(No.~17K05425) from Japan Society for the Promotion of Science.

% %%%%%%%%%%%%%%%%%%%%%%%%%%%%%%%%%%%%%%%%%%%%%%%%%%%%%%%%%%%%%%%%%%%%%%%%%%%%%%%%%%%%%%%%%%
% %%%%%%%%%%%%%%%%%%%%%%%%%%%%%%%%%%%%%%%%%%%%%%%%%%%%%%%%%%%%%%%%%%%%%%%%%%%%%%%%%%%%%%%%%%

\end{document}